\documentclass[aps, prb, preprint, lengthcheck, footinbib, showpacs, superscriptaddress, citeautoscript]{revtex4-1}
\usepackage[T1]{fontenc}
\usepackage[english]{babel}
\usepackage{amsmath}
\usepackage{braket}
\usepackage{graphicx}
\usepackage[colorlinks, urlcolor=blue, citecolor=blue, linkcolor=blue, pdfstartview=FitH]{hyperref}
\usepackage[all]{hypcap}

\begin{document}

\def\affiSOLAB{Spin Optics Laboratory, Saint~Petersburg State University, 198504 St.~Petersburg, Russia}
\def\affiRC{SPbU Resource Center ``Nanophotonics'', Saint~Petersburg State University, 198504 St.~Petersburg, Russia}
\def\affiSP{Departament of Physics, Saint~Petersburg State University, 198504 St.~Petersburg, Russia}
\def\affiSH{Physics and Astronomy School, University of Southampton, Highfield, Southampton, SO171BJ, UK}

\title{Nontrivial relaxation dynamics of excitons in high-quality InGaAs/GaAs quantum wells}
\author{A.~V.~Trifonov}
\author{S.~N.~Korotan}
\author{A.~S.~Kurdyubov}
\author{I.~Ya.~Gerlovin}
\author{I.~V.~Ignatiev}
\affiliation{\affiSOLAB}
\author{Yu.~P.~Efimov}
\author{S.~A.~Eliseev}
\author{V.~V.~Petrov}
\affiliation{\affiRC}
\author{Yu.~K.~Dolgikh}
\author{V.~V.~Ovsyankin}
\affiliation{\affiSP}
\author{A.~V.~Kavokin}
\affiliation{\affiSOLAB} \affiliation{\affiSH}

\date{\today }

\begin{abstract}
Photoluminescence (PL) and reflectivity spectra of a high-quality
InGaAs/GaAs quantum well structure reveal a series of ultra-narrow peaks
attributed to the quantum confined exciton states. The intensity of these
peaks decreases as a function of temperature, while the linewidths
demonstrate a complex and peculiar behavior. At low pumping the widths of
all peaks remain quite narrow ($< 0.1$~meV) in the whole temperature range
studied, 4 -- 30~K. At the stronger pumping, the linewidth first increases
and than drops down with the temperature rise. Pump-probe experiments show
two characteristic time scales in the exciton decay, $< 10$~ps and 15 --
45~ns, respectively. We interpret all these data by an interplay between the
exciton recombination within the light cone, the exciton relaxation from a
nonradiative reservoir to the light cone, and the thermal dissociation of
the nonradiative excitons. The broadening of the low energy exciton lines
is governed by the radiative recombination and scattering with reservoir
excitons while for the higher energy states the linewidths are also
dependent on the acoustic phonon relaxation processes.
\end{abstract}


\maketitle

\section*{Introduction}

Excitons in two-dimensional semiconductor structures with quantum wells
(QWs) and superlattices may be efficiently coupled to light because of the
breaking of the wave vector selection rules along the structure growth
direction. Radiative recombination rates of excitons strongly depend on
their localization radii. In ideal QWs, the radiative decay of an exciton is
possible only if the in-plane component of its wave vector does not exceed
one of the photon to be emitted, $K_{c}=n\omega /c$, where $n$ is the
refractive index, $\omega$ is the frequency of light. We note that 
$K_{c}=0.03$~nm$^{-1}$ at the exciton resonance frequency in GaAs QWs, 
which corresponds to an exciton kinetic energy: $E_{c}=\hbar
^{2}K_{c}^{2}/(2M_{X})=0.06$~meV, which is much lower than the
characteristic thermal energy of the system at the liquid helium
temperature. Here $M_{X}=0.5$ is the exciton mass in units of the free
electron mass. Excitons with larger wave vectors do not interact with light.
They will be referred to as nonradiative excitons. The oscillator strength
of the whole exciton branch is accumulated within its little part (light
cone) and the radiative decay rate can reach $10^{11}$ s$^{-1}$ in
GaAs-based structures with QWs~\cite{Rashba, Andreani-SSC1991,
Hanamura-PRB1988, Deveaud-PRL1991, Vinattieri-PRB1994, Koch-nmat2006}. 
For nonradiative excitons, the dominant mechanism of decay is the
phonon-mediated relaxation into the states with small wave vectors followed
by the radiative recombination.

An important role of the reservoir of nonradiative excitons was recently
recognized for semiconductor microcavities where strong light-matter
coupling accelerates the exchange between radiative and nonradiative 
states~\cite{Vishnevsky-PRB2012,Wouters-PRB2013,Giorgi-PRL2014,Demirchyan-PRB2014,Haug-PRB2014,Belykh-PRB2014}. 
In conventional QW structures, all the 
dynamic processes are slowed down compared to microcavities, that favors
considerable accumulation of nonradiative excitons. The effect of these
excitons on the radiative ones may be significant. Nonradiative excitons
can be created even at the strictly resonant excitation of the lowest
exciton transition if the phonon energy corresponding to the lattice
temperature is larger than the critical energy $E_{c}$~\cite{Piermorocchi-PRB1996}. 
The in-plane component of the phonon wave vector, $q_{\parallel }$, 
is transferred to the exciton due to the conservation of momentum 
that results in ejection of this exciton outside the light cone.

For excitons localized by structure defects as well as for excitons
scattered by free carriers, phonons and other excitons, the number of atomic
oscillators contributing to the exciton oscillator strength is reduced as
compared to the free exciton. The radiative decay rate for such excitons may
decrease, by up to two orders of magnitude~\cite{Feldmann-PRL1987,
Citrin-PRB1993, Kavokin-PRB1994}. Besides, structure imperfections may
induce additional relaxation processes for radiative as well as for
nonradiative excitons. In particular, the defect centers or the QW
interface roughness may give rise to the considerable inhomogeneous
broadening of exciton resonances and to the nonradiative exciton
recombination. For this reason, many attempts were done to study the exciton
dynamics in GaAs-based heterostructures of various sample quality.

The exciton dynamics was extensively studied using the photoluminescence
(PL) kinetics measurements. Under nonresonant excitation, the PL pulse is
characterized by a rise time, $\tau _{rise}=10-1000$~ps, and a decay time, 
$\tau _{decay}=1-30$~ns, depending on the experimental conditions as well as
on the sample design and quality~\cite{Feldmann-PRL1987, Damen-PRB1990,
Deveaud-PRL1991, Srinivas-PRB1992, Vinattieri-PRB1994, Szczytko-PRL2004,
Roussignol-PRB1992, Kappei-PRL2005, Deveaud-ChemPhys2005, Bajoni-PSS2006,
Portella-PRL2009}. The theoretical analysis of the data performed in these
works and also in Refs.~\cite{Piermorocchi-PRB1996, Hanamura-PRB1988,
Basu-PRB1992, Citrin-PRB1993, Ciuti-PRB1998, Ivanov-PRB1999} allowed
obtaining characteristic rates of processes occurring mainly in the
reservoir of nonradiative excitons. It was found that, the characteristic
time of exciton formation from free carriers lies in the range from several
tens to hundreds of picoseconds~\cite{Szczytko-PRL2004, Bajoni-PSS2006,
Portella-PRL2009, Piermarocchi-pss1997}. The exciton thermalization in the
reservoir occurs approximately in the same time 
range~\cite{Piermorocchi-PRB1996, Basu-PRB1992, Ivanov-PRB1999}. 
These two processes are mainly responsible for the PL rise time in high-quality 
structures. If radiative excitons are localized due to structure imperfections, 
their recombination time may become comparable with the time of exciton
thermalization in the nonradiative reservoir so that it affects the rise of
PL signal~\cite{Damen-PRB1990, Citrin-PRB1993, Kavokin-PRB1994}. The slowest
exciton dynamics process is the scattering of nonradiative excitons into
the light cone. This momentum relaxation process is responsible for the PL
decay. A large spread of the decay time~\cite{Feldmann-PRL1987,
Damen-PRB1990, Deveaud-PRL1991, Srinivas-PRB1992, Vinattieri-PRB1994,
Gurioli-PRB1991, Eccleston-PRB1992} is possibly caused by the competition of
the momentum relaxation with losses of nonradiative excitons via quenching
centers in real structures.

The PL kinetics experiments did not allow to direct measuring the
recombination time of radiative excitons in high-quality structures. The
first attempts to measure this time at the strictly resonant excitation
described in Refs.~\cite{Deveaud-PRL1991, Vinattieri-PRB1994,
Eccleston-PRB1992, Filkenstein-PRB1998} were not very successful because of
the limited time resolution of the setups used. The recombination as well as
the dephasing time for radiative excitons was extensively studied in the
pump-probe and four-wave mixing experiments~\cite{Schultheis-PRB1986,
Honold-PRB1989, Kim-PRL1992, Duer-PRL1997, Borri-PRB1999, Chemla-Nature2001,
Smith-PRL2010}. Results of these studies also reveal large variations of the
exciton recombination time in the range 1 --- 30~ps, most probably due to
the different quality of the investigated structures.

The large spread of experimental data on the exciton relaxation times in
structures with QWs points out that the reliable data on characteristic
rates of the relaxation processes both for radiative and for nonradiative
excitons can be obtained only by the careful selection of high-quality QW
structures and in specific experimental conditions.

In this paper, we report on the experimental study of the dynamics of
radiative and nonradiative exciton states in a specially designed
high-quality heterostructure with a relatively wide InGaAs/GaAs QW of about
95~nm width. The peculiarity of this structure is in the presence of a set
of the quantum confined exciton states, which manifest themselves as the
ultra-narrow lines in the PL and reflection spectra. The PL spectra and the
kinetics of pump-probe signal were measured under resonant excitation into
one of the quantum confined states that allowed us to carefully control the
excitation conditions. A comparative analysis of PL and reflectance spectra 
and of kinetics data has been performed for
reliable identification of relaxation processes and extraction of their
parameters. In particular, our analysis has shown that the dominant decay
process for excitons at the lowest energy level is the radiative
recombination for the case of low temperature and low excitation power.

We have found that the kinetics of pump-probe signal for the lowest exciton
state exhibits, besides the fast component corresponding to the radiative
exciton decay, a slow component, whose decay time is longer by three orders
of magnitude. The long-lived component of the kinetics is caused by the
relaxation of nonradiative excitons from the reservoir to the light cone.
We have also identified other relaxation processes in the exciton system
under study. In particular, we have found that the increase of excitation
density is followed by the enhancement of the exciton-exciton and/or
exciton-carrier collisions which broaden exciton lines in the PL spectra. At
the same time, the temperature-activated rapid quenching of the PL intensity
does not accompanied by any noticeable broadening of exciton lines. We
attribute this unusual effect to the thermal dissociation of nonradiative
excitons in the reservoir followed by the nonradiative losses of the
excitation energy.

\section{Experimental details}
\label{sect:1}

We have studied an InGaAs/GaAs quantum well (QW) heterostructure grown by a
molecular beam epitaxy at an n-doped GaAs substrate with orientation [001].
The structure was grown at elevated temperature of the substrate of about
550~$^{\circ }$C to prevent clusterization of Indium atoms~\cite{Uddin-JAP1989}. 
The structure contains a wide InGaAs QW layer with a
nominal thickness of about 95~nm and the Indium content of 2\%. Besides, it
contains a reference narrow QW ($L=2$~nm) with Indium content of 2\% also.
There is a gradient of the layer thicknesses of about 10\% per cm and the
Indium content varies from 1.5 to 2 \% in the structure so that the real
thickness and potential profile of the wide QW were estimated from
spectroscopic data. The sample was cooled using an optical
cryostat with a close cycle of helium cooling. The sample temperature was
varied in the range of 4 -- 30 K.

The PL was excited by radiation of a tunable continuous-wave Ti:sapphire
laser. We excited the QW excitons either quasi-resonantly into one of the
quantum confined exciton levels or higher in energy up to the absorption edge
of GaAs barriers. The PL was dispersed and detected by a 0.55 m spectrometer
equipped with a CCD. The spectral resolution in actual range was about 
30~$\mu$eV. Laser spot on the sample was about 50~$\mu$m and excitation 
power varied in range 2 -- 150~$\mu$W. The PL excitation (PLE) spectra were
measured using the same setup by means of continuous tuning of the laser
wavelength and detecting the PL spectra in the range of several lowest
exciton transitions for each excitation wavelength. We would like to note
that, due to high quality of the sample, the intensity of PL from two lowest
exciton levels dominated over the scattered light of laser tuned to photon
energy only one meV above these levels.

Another experimental technique exploited in our study is the pump-probe
method. We used pulsed radiation of a femtosecond Ti:sapphire laser, split
into the pump and probe beams. The pump beam passed through an
acousto-optical tunable filter, which cut out a spectrally narrow pulses
with full width at half maximum of about 0.6~nm corresponding to
spectrally-limited pulse duration, $T_{pu} = 1.75$~ps. The pump wavelength 
was tuned to excite predominantly the lowest one or two exciton transitions
only. The probe pulses were spectrally broad and short, of about 0.1~ps. To
detect the time-resolved photomodulated reflection at different exciton
transitions, the spectrum of reflected probe beam was analyzed using a 0.55
m spectrometer equipped with a photodiode.

\section{Experimental results}

\subsection{PL, PLE, and reflection spectra at low temperature}

\label{sect:3.1}

\begin{figure}[ht]
\includegraphics[width=1\columnwidth]{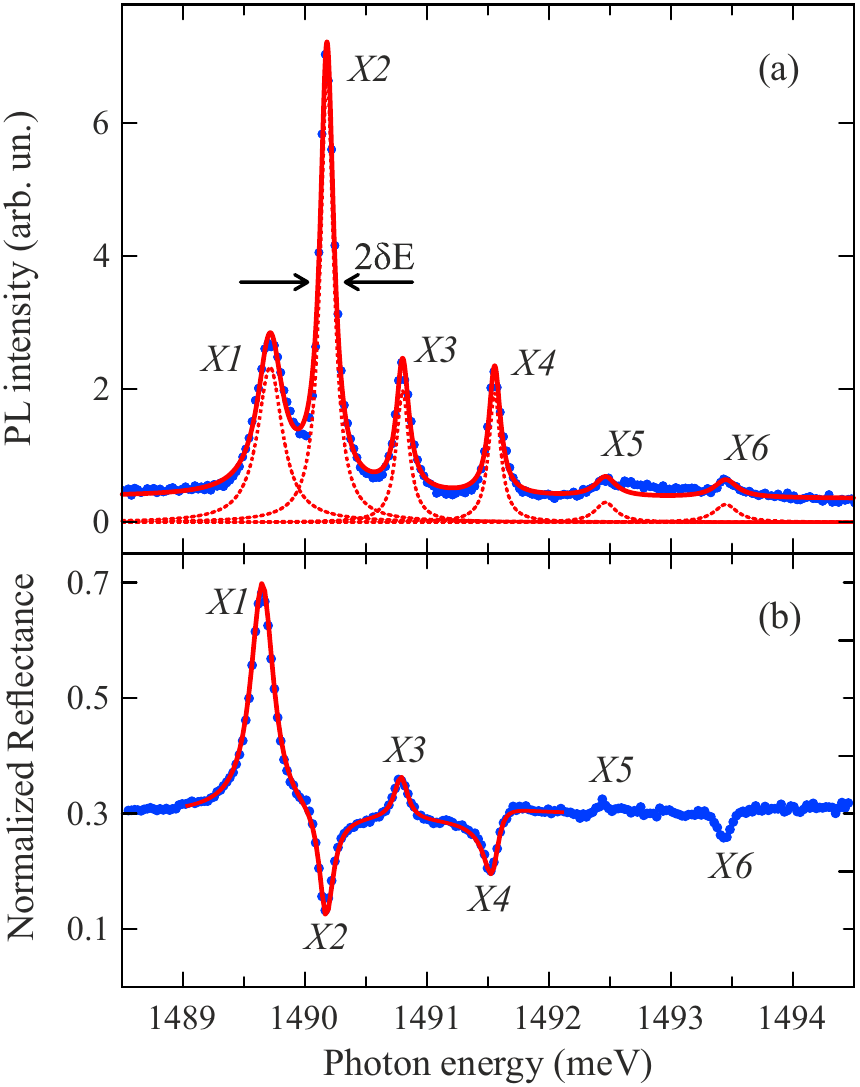}
\caption{(Color online) (a) PL spectrum of the InGaAs/GaAs QW with nominal width 
$L=95$~nm measured at temperature $T = 4$~K and excitation power 
$P=50$~$\protect\mu$W (blue dots). The PL is excited into the $X8$ exciton
level. Thin dashed curves are the Lorentzians 
with parameters listed in Tab.~\ref{table1} for the first four transitions. 
Notations "XN'' indicate transitions from the quantum confined exciton levels.
The sum of Lorentzians is shown by red solid curve. 
(b) Reflectance spectrum normalized to the excitation
intensity (blue dots). Red solid curve shows the spectrum calculated 
in framework of the model described in the text with parameters listed 
in Tab.~\ref{table1}.}
\label{fig1}
\end{figure}

The PL and reflectance spectra taken from the same spot on the sample are
shown in Fig.~\ref{fig1}. The spectra consist of several narrow peaks
(resonances), whose positions and widths coincide in both spectra with high
accuracy. The absence of any detectable Stokes shift (within experimental
error in a few tens of $\mu $eV) between the peaks in the PL and reflectance
spectra indicates high quality of the sample. 

It is naturally to assume that the observed resonances are due to quantization
of exciton motion across the QW layer. In this case, the energies of resonances can
be roughly estimated using a simple model of a wide QW with infinitely high
barriers, in which the exciton is quantized as a whole~\cite{Ivchenko-book}:
\begin{equation}
E_{XN} = E_g - E_B +\frac{\hbar}{2M_X}\left(\frac{\pi N}{L^*}\right)^2,
\label{quantization}
\end{equation}
where $E_g$ is the band gap, $E_B$ is the exciton binding energy (exciton
Rydberg), $M_X$ is the exciton mass, and $L^*$ is the effective thickness of
QW layer, which is smaller than the real width $L_{QW}$ by the dead
layer width, $\delta L_d$~\cite{Loginov, Schiumarini-PRB2010}. The
dead layer width is governed by the exciton Bohr radius, $a_B$. In 
our case, $L^* = L_{QW} - 2 \delta L_d \approx 75$~nm assuming 
that $\delta L_d = 0.8 a_B = 10$~nm as it is estimated for QWs with 
a layer thickness of about $6 a_B$~\cite{Schiumarini-PRB2010}. Using 
the above equation, we obtain the following energy positions of quantum 
confined states relative to the lowest one: 
$\left(E_{XN} - E_{X1}\right)_{calc} = 0$, 0.40, 1.07, 2.00 (in meV) 
for $N = 1 \ldots 4$.   
The obtained values well agree with the energy distances between 
the {\it XN} and {\it X1} peaks found in the experiments (see Fig.~\ref{fig1} and 
Tab.~\ref{table1}): $\left(E_{XN} - E_{X1}\right)_{exp} = 0$, 0.50, 1.12, 1.88 
(in meV) for $N = 1 \ldots 4$.

The resonances in the PL spectrum can be well fitted by a set of Lorentzians, 
$L_{XN}(E)=(A_{XN}/\pi )\delta E_{XN}/[(E-E_{XN})^{2}+\delta E_{XN}^{2}]$,
superimposed on a background signal. Here $\delta E_{XN}$ is
the half width at half maximum (HWHM) of the Lorentzian and $A_{XN}$ is the
area under the Lorentzian curve. The parameters of Lorentzians are given in 
Tab.~\ref{table1}.
Because of gradients of the QW width and of the Indium content, 
the relative PL intensities from different exciton levels, $A_{XN}/A_{X1}$, 
slightly depends on the point on the structure surface. Therefore most 
of the measurements were done at the same spot of the sample.
It should be emphasized that the PL intensity for higher exciton states is
comparable with that for the lowest energy state. This feature strongly
distinguishes the QW structure under study from those studied by other
authors, where the high-energy exciton peaks in PL spectra are virtually 
absent (see, e.g., Ref.~\cite{TredicuchiPRB1993}).

The analysis of the PL spectrum does not provide a direct knowledge of the
radiation rates because the intensity of a PL peak depends also on the
exciton population, which can be different for different quantum confined
states. As one can see from Tab.~\ref{table1}, the integral intensities of the
first and second exciton peaks in the PL spectrum are very similar, but this
indicates only that the population of the second exciton level is
significantly larger than that of the first one.

In a contrast to the PL spectrum, the quantitative analysis of reflectance 
spectrum allows one to obtain valuable information about radiative and 
nonradiative decay rates for different exciton states. First, we would like to note 
that the maximum reflectivity at the first resonance is about 0.7, not far from unity
corresponding to the totally ``metallic'' reflection. This is the first indication 
that the radiative decay of these excitons prevails over the nonradiative one.

We have performed a simple analysis of the reflectance spectrum 
generalizing the theory developed in Ref.~\cite{Ivchenko-book} for 
the case of several exciton quantum confined states. The coefficient 
of amplitude reflectance from a QW with several exciton resonances 
can be written in the form:
\begin{equation}
r_{QW} = \sum\limits_{N=1}^{N_{max}} \frac{i(-1)^{N-1}\Gamma_{0N}e^{i\varphi_N}}%
{\omega_{0N} - \omega - i(\Gamma_{0N}+\Gamma_N)}.
\label{IvchenkoEq}
\end{equation}
Here $\omega_{0N}$ is the resonance frequency, $\Gamma_{0N}$ 
and $\Gamma_N$ are the radiative and nonradiative damping rates. 
Phase $\varphi_N$ appears in this equation due to an asymmetry of the QW potential 
caused by the Indium segregation during the heterostructure growth 
process~\cite{Muraki-APL1992}. Reflectance, $R(\omega)$, 
from the structure with a top barrier layer of thickness $L_b$ and a 
QW layer of thickness $L_{QW}$ is calculated by a standard 
manner~\cite{Ivchenko-book}:
\begin{equation}
R(\omega) = \left|\frac{r_{01}+r_{QW}e^{2i\phi}}{1+r_{01}r_{QW}e^{2i\phi}}\right|^2,
\label{Reflectance}
\end{equation}
where $r_{01}$ is the amplitude reflectance from the sample surface. 
Phase $\phi = K(L_b + L_{QW}/2)$ where $K$ is the photon wave vector 
in the heterostructure. 

Result of simulation of the reflectance spectrum using Eqs.~(\ref{IvchenkoEq}), (\ref{Reflectance}) 
is shown in Fig.~1(b). Respective fitting parameters are listed in Tab.~\ref{table1}. 
As seen the calculated curve very well reproduces the experimentally measured 
spectrum. Although the number of fitting parameters is large, they are well 
defined because determine different peculiarities of the resonances. 
In particular, ratio $\Gamma_{0N}^2/(\Gamma_{0N}+\Gamma_{N})^2$ determines 
the peak amplitude of resonance $N$ and $\hbar(\Gamma_{0N}+\Gamma_{N})$ determines 
its HWHM. Phase $\varphi_N$ determines the asymmetry of resonance that is its 
deviation from a Lorentzian, which is clearly seen, e.g., for the $X4$ resonance 
in Fig.~\ref{fig1}(b). 

\begin{table}[h]
\caption{\label{table1}Fitting parameters of the PL and reflectance spectra. }
\begin{center}
\begin{tabular}{|c|c|c|c|c|c|}
\hline
 Exp. &Exciton level & X1 & X2 & X3 & X4 \\
\hline
PL & A (arb. un.) & 1080 & 1390 & 390 & 420 \\  
\hline
PL & E (meV) & 1489.71 & 1490.18 & 1490.80 & 1491.56 \\
\hline
PL & $\delta E$ ($\mu$eV) & 123 & 67 & 63 & 58 \\
\hline
Refl & $\hbar \omega_0$ (meV) & 1489.65 & 1490.17 & 1490.79 & 1491.55 \\
\hline
Refl & $\hbar\Gamma_0$ ($\mu$eV) & 47.5 & 19.5 & 6 & 11 \\
\hline
Refl & $\hbar\Gamma$ ($\mu$eV) & 38.5 & 60 & 57 & 66 \\
\hline
Refl & $\varphi$ (rad.) & 0 & -0.06 & -0.03 & 0.40 \\
\hline
\end{tabular}
\end{center}
\end{table}

The comparison of the obtained radiative and nonradiative damping rates shows 
that the radiative broadening is really prevail for the first exciton resonance. 
Real magnitudes of nonradiative broadening are even smaller than those given in 
Tab.~\ref{table1} because the limited resolution of our setup (of about 30~$\mu$eV) 
gives rise to an additional broadening the resonances measured.
We should note that  the spectral broadening of the lowest transition in both the PL 
and reflection spectra is of about 0.1~meV that is extremely small for structures of this type 
known in literature, see, e.g., 
Refs~\cite{Srinivas-PRB1992, Eccleston-PRB1992, Poltavtsev, Chen87, Singh-PRB13}.

We estimate the ratio of radiative decay rates of different exciton
transitions using a simplified model of size quantization of an exciton motion in a
symmetric QW with infinitely high barriers~\cite{Tuffigo}. The envelope wave functions
of the center-of-mass motion of the exciton as a whole are proportional to 
$\cos (N\pi z/L^{\ast })$ for odd number $N$ of the quantum confined exciton
state and to $\sin (N\pi z/L^{\ast })$ for even $N$, where $z$ is the exciton center
of mass coordinate along the structure axis. 

The rate of the radiative transition, $\Gamma _{0N}$, is proportional to
the squared overlap integral of the exciton wave function with the light
wave, $F_{N}^{2}(K)$, which, in framework of this model, 
is~\cite{Tuffigo}:
\begin{equation}
F_{N}^{2}\left( K\right) =%
\begin{cases}
C_{N}^{2}\cos ^{2}{\left( K L^{\ast }/2\right) }\text{~for~odd~}N, \\ 
C_{N}^{2}\sin ^{2}{\left( K L^{\ast }/2\right) }\text{~for~even~}N,
\end{cases}
\label{eq:Tuff1}
\end{equation}
where 
\begin{equation}
C_{N}=\frac{N\pi /L^{\ast }}{\left[ \left( N\pi /L^{\ast }\right)
^{2}-K^{2}\right] }\sqrt{2/L^{\ast }}.  
\label{eq:Tuff2}
\end{equation}
Here $K$ is the photon wave vector in the QW material. As seen from
Eqs.~(\ref{eq:Tuff1}, \ref{eq:Tuff2}), this rate may be strongly different
for exciton states with even and odd numbers and gradually decreases with 
$N$. For the case of QW under study, equations (\ref{eq:Tuff1}) and (\ref{eq:Tuff2}) 
give the following radiative decay rates for different quantum confined states 
normalized to the rate of $X1$ transition:
$(\Gamma_{0N}/\Gamma_{01})_{calc}= F^2_N(K)/F^2_1(K) = 1$, 0.28, 0.04, 0.06 
for $N = 1, \ldots 4$. 
Numerical calculations for QWs of similar widths in more accurate models give rise 
to similar results~\cite{Schiumarini-PRB2010}. The dependence of normalized rate on 
number $N$ of exciton state qualitatively reproduces the dependence obtained 
from the reflectance spectrum: 
$(\Gamma_{0N}/\Gamma_{01})_{exp} = 1$, 0.41, 0.13, 0.23. At the same time, 
the quantitative disagreement between the calculated and experimental data indicates 
that the exciton wave functions in real structure considerably differ from those 
used in the above simple model. 
 
The exciton radiative time can be estimated from the given in Tab.~\ref{table1} 
values of $\Gamma_{0N}$ using relation: $\tau _{XN}= 1/(2\Gamma_{0N})$, 
see, e.g., Ref.~\cite{Ivchenko-book}, p. 92. For the first two transitions, 
the radiative times are: $\tau _{X1} = 6.9$~ps and $\tau _{X2} = 17$~ps. 
These values are of the same order of magnitude as the radiative
recombination time, $\tau _{rad}\sim 10$~ps, reported previously for GaAs 
QWs~\cite{Deveaud-PRL1991, Vinattieri-PRB1994}.
As it seen from Tab.~\ref{table1}, the nonradiative broadening $\hbar\Gamma_{X1}$ 
for the first exciton state is smaller than for the higher states.
This means that the broadening of excited states is contributed by additional 
nonradiative mechanisms.

\begin{figure}[t]
\includegraphics[width=1\columnwidth]{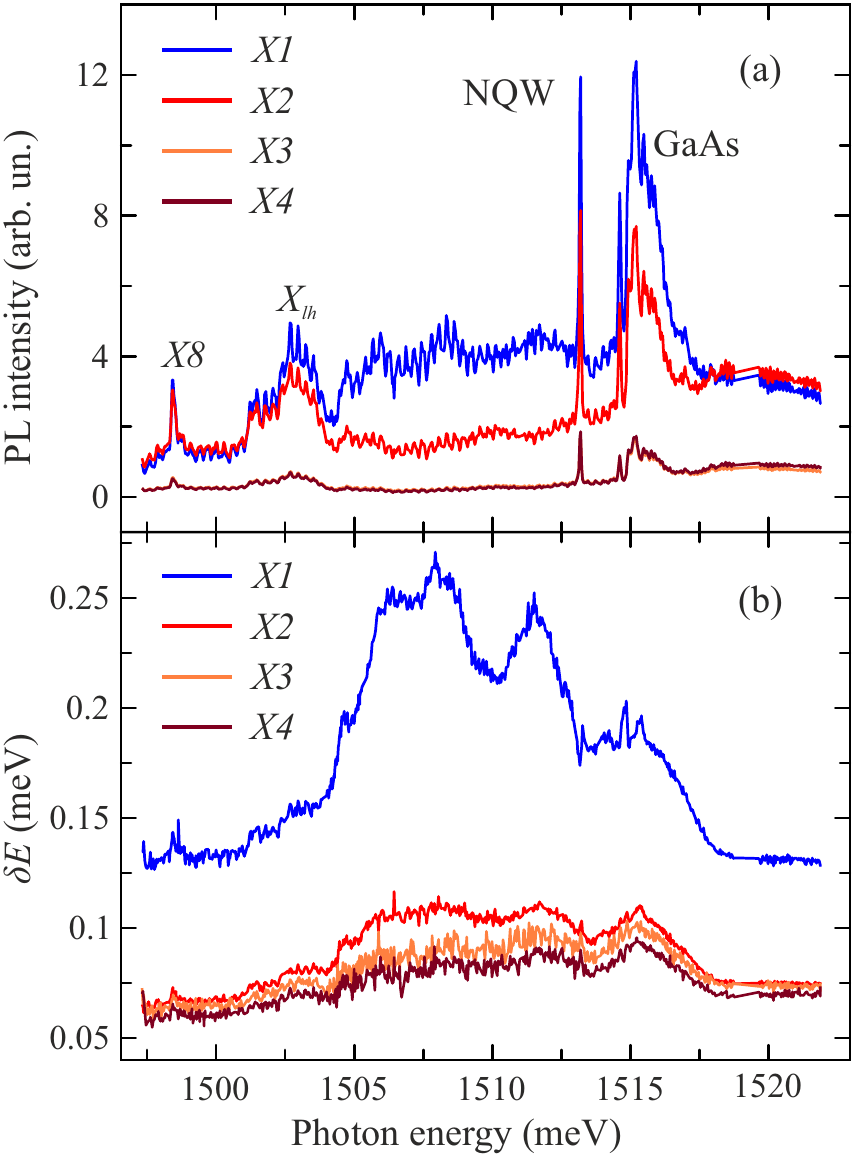}
\caption{(Color online) PLE spectra for resonances $X1, \ldots X4$. Legends
show identification of separate maxima: $X8$ is the eighth quantum confined state
of the heavy-hole exciton; $X_{lh}$ is the lowest state of the
light-hole exciton; ``NQW'' is the lowest state of the heavy-hole exciton in
the narrow QW; ``GaAs'' is the exciton state in the GaAs barriers. (b) HWHM of the
resonances as functions of excitation energy. Excitation power 
$P=50$~$\protect\mu$W. }
\label{fig2}
\end{figure}

For a deeper insight into the origin of relaxation processes in the
structure, we have measured the PLE spectra of the exciton resonances. 
Fig.~\ref{fig2}(a) shows the spectral dependences of integral intensity for the
first four peaks on the photon energy of excitation. The spectra display several 
remarkable features~\cite{footnote}. The $X_{lh}$ band was identified as 
the lowest state of the light hole exciton. The degree of circular polarization 
of PL measured at this resonance has the opposite sign compared to the 
polarization degree of the heavy hole resonances. This effect is due to different 
selection rules for respective optical transitions and is well known in literature~%
\cite{NCP}. The $X_{lh}$ exciton state is split off from the $X1$ exciton mainly 
due to the internal strain in the GaAs/InGaAs structure caused by the lattice
mismatch of the QW and barriers.

The efficiency of excitation of all the resonances is synchronously changed
with the photon energy increase up to the energy of the transition $X_{lh}$.
However, above this transition, the lowest exciton state ($X1$) is populated
more efficiently than other states, so that the relative intensity of the
corresponding exciton resonance increases. This is an indication that one
more relaxation mechanism is ``switched on'' in this spectral range of 
excitation. Its possible origin will be discussed in Section \ref{sect:disc}.

Fig.~\ref{fig2}(b) shows the dependence of HWHM of the first four exciton
peaks on the excitation photon energy. The HWHM were obtained by Lorentzian
fitting of the PL spectra measured at each excitation energy. The excitation
energy was scanned with small step of about 0.05~meV. As one can see from
the figure, the peak broadening noticeably increases above the $X_{lh}$
transition. A particularly strong increase of HWHM is observed for the first
resonance that also indicates an additional mechanism of the 
broadening, which switches on at these photon excitation energies.

\subsection{Temperature and pump power dependences of the exciton resonances}

\label{sect:temp-power}

\begin{figure}[t]
\includegraphics[width=1\columnwidth]{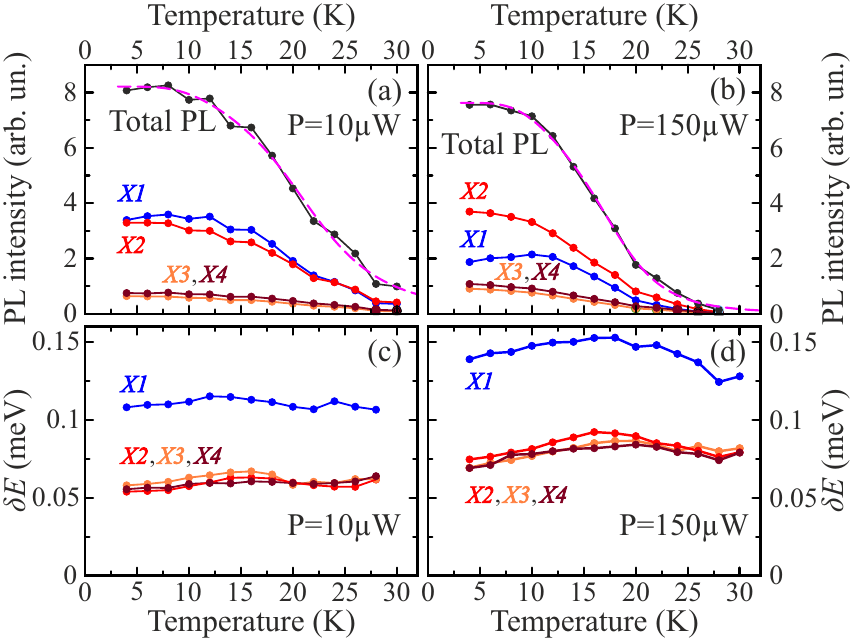}
\caption{(Color online) Temperature dependencies of integral PL intensities
(a, b) and of HWHM for different resonances (c, d). Solid curves in (a) and (b)
are the fits by Eq.~(\protect\ref{eq:Arr}) with parameters: 
$\protect\gamma_{0X}/\protect\gamma_r = 1.1$, 
$\protect\gamma_{0b}/\protect\gamma_r=780$ for excitation power 
$P=10$~$\protect\mu$W in (a); 
$\protect\gamma_{0X}/\protect\gamma_r = 0.4$, 
$\protect\gamma_{0b}/\protect\gamma_r=80$ for excitation power 
$P=150$~$\protect\mu$W in (b). The activation energies for both
excitation powers are: $E_x = 4.5$~meV and $E_b = 16$~meV. The dependences
were measured at the excitation into the $X6$ transition, whose spectral
position was determined at each temperature. }
\label{fig3}
\end{figure}

The experimental data discussed in the previous section have been obtained
at low temperature (4~K) and low excitation power (10 -- 50~$\mu $W). In
this case the phonon-mediated relaxation and the exciton-exciton
scattering are not very efficient. To investigate the role of
phonon-mediated processes, we have studied the temperature variations of PL
spectra. The temperature rise has been found to be accompanied by a
synchronous decrease of the integral amplitude of all exciton resonances,
leading to the approximately 20-fold decrease of the total PL intensity with
the temperature increase from 4 to 30~K (see Fig.~\ref{fig3}). The
temperature dependence of the total PL intensity can be well approximated by
expression: 
\begin{equation}
I\left( T\right) =\frac{P }{1+\left(\gamma _{0X}/\gamma_r\right)%
\exp \left( -\frac{E_{X}}{kT}\right) +\left(\gamma _{0b}/\gamma_r\right)%
\exp \left( -\frac{E_{b}}{kT}\right) }.
\label{eq:Arr}
\end{equation}
This expression is derived from the balance equation for the exciton
population $n_{X}$ accounting for the relaxation to radiative states with rate $\gamma
_{r}$(T) as well as for two processes of thermally-activated
dissipation of excitons. In such conditions, the balance equation reads: 
\begin{equation}
\frac{dn_{X}}{dt}=P-\left[ \gamma _{r}(T)+\gamma _{X}(T)+\gamma _{b}(T)\right]%
n_{X}.  
\label{eqn:7}
\end{equation}
Here $P$ is the rate of optical excitation. A theoretical analysis 
shows~\cite{Andreani-SSC1991, Feldmann-PRL1987, Rosales-PRB2013} that
the exciton relaxation time, $1/\gamma _{r}(T)$,  in QWs is proportional 
to the sample temperature. The temperature dependences
of the exciton dissipation rates are described by Boltzmann functions: 
\begin{equation}
\gamma _{i}(T)=\gamma _{0i}\exp (-E_{i}/kT).  
\label{eqn:8}
\end{equation}
The value of the first activation energy obtained in the fit, $E_{X}=4.5$~meV, 
approximately corresponds to the exciton binding energy. Therefore, we
assume that the first temperature-activated process is the exciton
dissociation into free carriers. The second energy, $E_{b}=16$~meV, is
significantly smaller than the quantization energy for excitons in the
QW under study (of about 25~meV). It appears that the obtained value of 
$E_{b}$ is closer to the band offsets for free electrons and/or holes
although the ratio of latter quantities is unknown and has been extensively
discussed in literature up to now, see, e.g., 
Refs~\cite{JoycePRB91,ZubkovPRB04,Biswas-Materials14}. This is why we conclude 
that the second temperature-activated process of PL quenching is the emission 
of carriers into barrier layers accompanied, possibly, by a radiative
recombination in a different spectral range.

Besides the PL quenching, the temperature increase helps establishing
thermal equilibrium between occupations of different exciton states. In this
case, we could expect a remarkable change of the relative intensities of
exciton peaks with temperature. In particular, at low temperatures, when the
thermal energy, $kT$, is smaller than the energy distance between the
quantum confined exciton states, the lowest exciton state should be predominantly
occupied. At elevated temperatures, when $kT>E_{X4}-E_{X1}$, the populations
of these levels should be nearly equal. The experimental data, however, show
that the relative intensities of different PL lines are almost independent
of temperature. This means that the efficiency of thermally activated
exciton dissociation and carrier ejection is higher that the phonon-mediated
transitions between different exciton levels responsible for the thermal
equilibrium. We conclude that the thermal equilibrium in the exciton
sub-system is not achieved in this structure in the temperature range of 
4 -- 30 K.

The widths of exciton lines also reveal a surprising, at first glance,
behavior at elevated temperatures. A strong decrease of the PL yield with
the temperature increase indicates that the nonradiative exciton relaxation
begins to dominate over the radiative one at $T>15$~K, see Fig.~\ref{fig3}(a, b). 
If the widths of exciton lines are controlled at low temperature
mainly by the radiative processes, the temperature rise should result in
line broadening, which is not observed in the experiment. Indeed, as seen in
Fig.~\ref{fig3}(c), the line widths measured at the low excitation power are
almost constant in the temperature range studied within our experimental
accuracy.

The increase of the excitation power stronger affects the line widths.
Figure~4 shows that it causes an additional line broadening, which is
approximately proportional to the square root of the pump power. The
additional broadening is a nonmonotonic function of temperature: it
increases with the temperature rise up to, approximately, $T=15$~K and then
falls down with the further sample heating [comp. Figs~\ref{fig3}(c) and
(d)]. This behavior is observed for all the excitation powers we used where
the additional broadening can be reliably identified.

\begin{figure}[t]
\includegraphics[width=1\columnwidth]{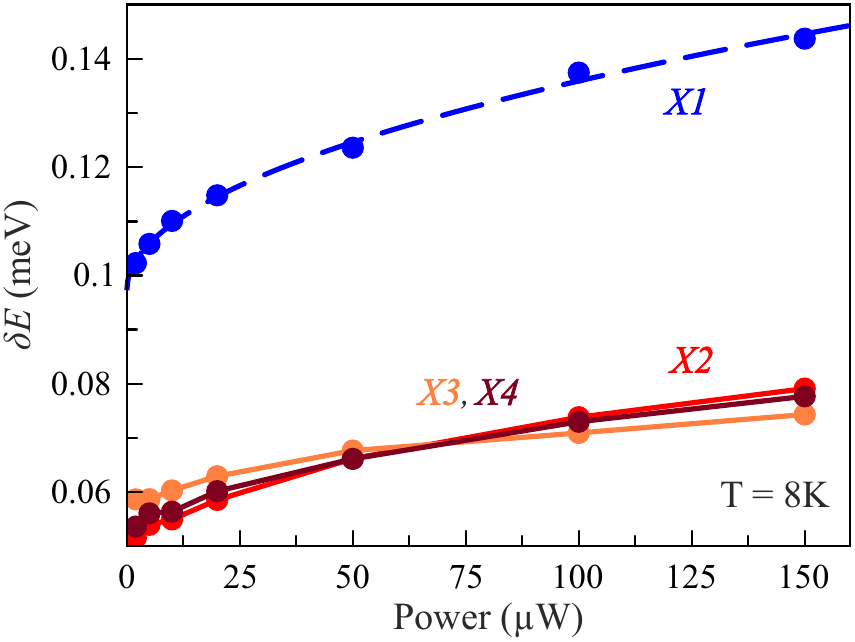}
\caption{(Color online) Power dependences of HWHM of the resonances measured
at $T = 8$~K (symbols). Curves are guides for the eye. }
\label{fig4}
\end{figure}

\subsection{Kinetics of exciton states}

\begin{figure}[t]
\includegraphics[width=1\columnwidth]{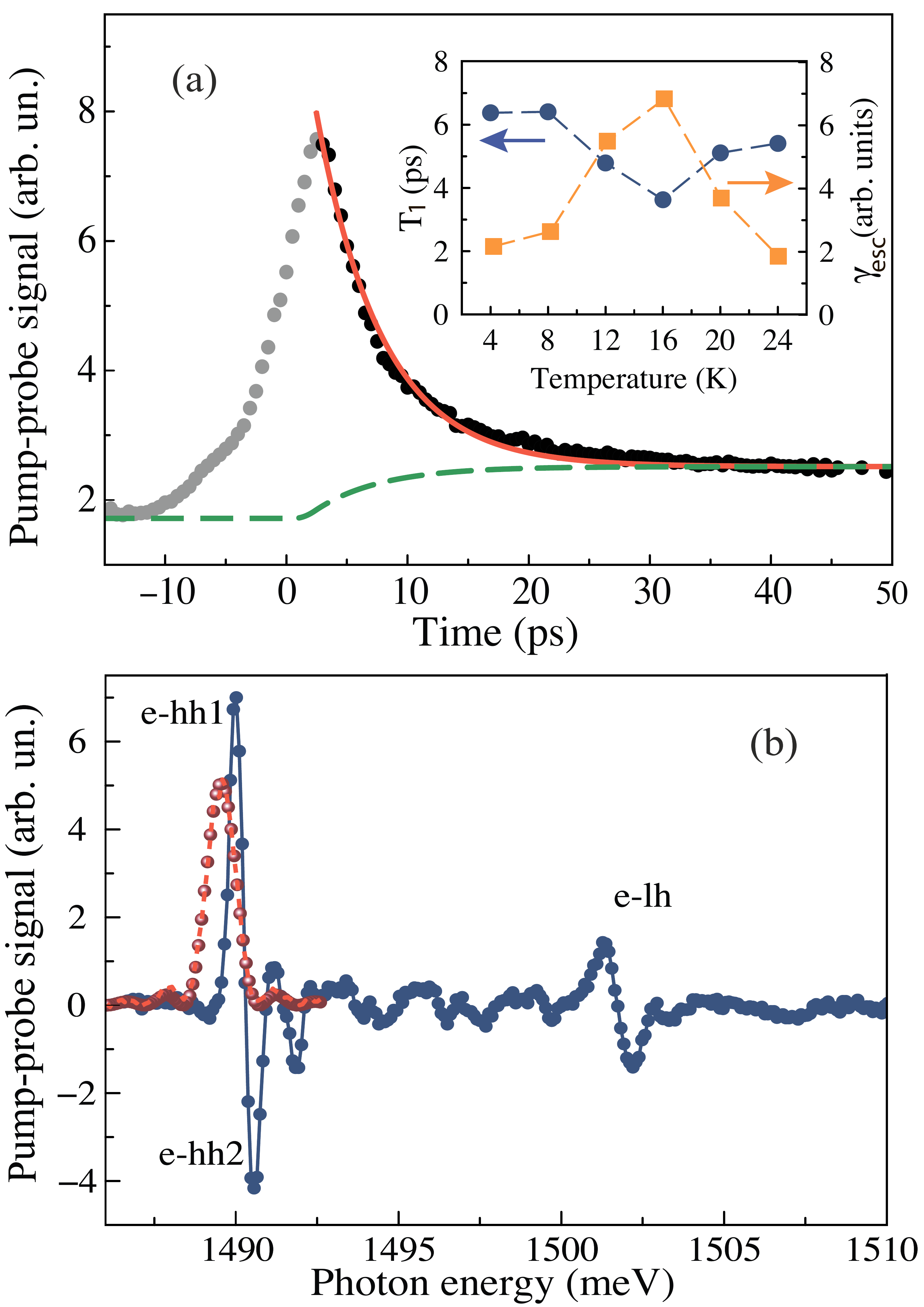}
\caption{(Color online) (a) Kinetics of pump-probe signal measured at the $X1$ 
resonance at $T = 4$~K and pump power $P_{pu} = 10$~$\mu$W (dots). 
The smooth solid curve shows the fit of the fast component decay 
by function $f_{fast}(t)$ [see Eq.~(\protect\ref{kin3})]. The dashed curve shows
the rise of slow component modeled by Eq.~(\protect\ref{kin2}).
Inset shows temperature dependences of $T_1$ (blue dots) and 
$\protect\gamma_{esc}$ (orange squares). (b) Spectral dependence of
differential reflection measured at the 30-ps delay after the pump pulse (blue curve).
Dotted red curve shows the spectrum of the pump pulse. Dashed curve is a
modeling of the pulse spectrum by function $I_{pu}(\protect\omega)$ given by 
Eq.~(\protect\ref{pump-spectrum}) with 
$\hbar \protect\delta \protect\omega = 0.37$~meV.} 
\label{fig5}
\end{figure}

The most direct knowledge of the exciton relaxation processes can be
obtained in kinetic experiments. We have performed pump-probe experiments in
the reflection geometry. Figure \ref{fig5}(a) shows the differential
reflectance signal for the lowest exciton transition as a function of delay
between pump and probe pulses. An important peculiarity of the signal is 
the simultaneous presence of fast and slow components. 
The slow component is almost unchanged during a few hundreds of picoseconds.
The fast component of kinetics reflects the dynamics of radiative excitons. In general case, 
it can be formed by the coherent and incoherent contributions. The coherent part 
of the signal is due to the four-wave mixing process~\cite{Shah-book}. 
It forms the leading edge of fast component when the probe pulse 
precedes the pump one, $t_{pr} < t_{pu}$. The analysis of coherent signal in detail is 
out of scope of present work.

The falling edge of fast component is formed by an ordinary pump-probe signal, 
which is determined by the exciton population created by the pump pulse. 
Assuming that the pump pulse has a finite duration  ($T_{pu} = 1.75$~ps, 
see Sect.~\ref{sect:1}), and the probe pulse is very short  ($\delta$-pulse), 
we obtain following expression for the decaying part of signal:
\begin{equation}
f_{pp}(t_{pr}) = I_{pr}\int I_{pu}(t)\theta(t_{pr}-t)%
\exp\left(-\frac{t_{pr}-t}{T_1}\right)dt. 
\label{kin1}
\end{equation}
Here we integrate over pump pulse and take into account the sequence 
of pulses introducing the step-like $\theta$-function. Quantities 
$I_{pu}(t)$ and $I_{pr}$ are the intensities of the pump and probe beams.
The pump-probe signal is proportional to the first powers of the pump and 
probe intensities and its dependence on the time delay is determined by 
the decay time of exciton population, $T_1$.

The fast component is superimposed on the slow one caused by 
the nonradiative excitons. It arises due to the ejection of radiative excitons 
beyond the light cone. 
Therefore the rise of this component is also fast and modeled as:
\begin{equation}
f_{slow}(t_{pr})= \alpha_{nr} \gamma_{esc}\int_{0}^{t_{pr}}f_{pp}(t)dt,
\label{kin2}
\end{equation}
where $\gamma_{esc}$ is the rate of exciton ejection and constant $\alpha_{nr} \le 1$ 
takes into account that the contribution of nonradiative excitons into the pump-probe
signal may be smaller than that of radiative ones for the same exciton density.

The total signal detected in the experiment is the sum of both the contributions:
\begin{equation}
f_{fast}(t_{pr}) = f_{pp}(t_{pr}) + f_{slow}(t_{pr}).
\label{kin3}
\end{equation}
In the modeling of pump-probe signal, we assume that the pump pulse has 
a rectangular temporal profile with duration $T_{pu}$. Its spectrum shown
in Fig.~\ref{fig5}(b) can be well fitted by function:
\begin{equation}
I_{pu}(\omega) = I_{pu0} \left(\frac{\sin[(\omega - \omega_0)/\delta\omega]}%
{(\omega - \omega_0)/\delta\omega}\right)^2.
\label{pump-spectrum} 
\end{equation}
This function is obtained by the Fourier transformation of a rectangular optical
pulse with carrier optical frequency $\omega_0$ and duration 
$T_{pu} = 1/\delta\omega$. 

Figure~\ref{fig5}(a) shows the fit of falling edge of the fast component by function 
$f_{fast}(t)$. The rise of slow component is also shown by a separate curve. 
One can see that the decay of fast component is well modeled by function $f_{fast}(t)$ 
that allows us to obtain decay time $T_1$ of exciton population. For temperature 
$T = 4$~K, time $T_1 = 6.4$~ps, which is close to that obtained from 
the reflectance spectrum, $1/\Gamma_{01} = 6.9$~ps.   

The inset in this figure shows the temperature dependence of $T_1$. As seen, 
it strongly decreases when the temperature increases up to 16~K and then rises
again. To analyze this effect, we assume that the decay of fast component
is determined by the two main processes, the radiative decay 
and the exciton ejection characterized by rate $\gamma_{esc}$. 
Correspondingly, the decay rate is:
\begin{equation}
 1/T_1 = \Gamma_{01} + \gamma_{esc}.
 \label{kin4}
\end{equation}
The temperature dependence of $\gamma_{esc}$ can be obtained from 
experimental data. Indeed, the amplitude of slow component at time $t >> T_{pu}$
becomes independent of time and is determined by $\gamma_{esc}$. 
From Eq.~(\ref{kin2})] we obtain:
 \begin{equation}
 \Delta f_{slow} = \frac{f_{slow}}{f_{pp}(T_{pu})}%
  = \alpha_{nr} \gamma_{esc} T_1 e^{T_{pu}/T_1}.
 \label{kin5}
\end{equation}
Here we normalized the amplitude of slow component on the peak amplitude 
of pump-probe signal described by Eq.~(\ref{kin1}) to exclude intensities 
of the pump and probe beams.
The temperature dependence of $\gamma_{esc}$ is shown in the
inset of Fig.~\ref{fig5}(a). As seen this rate strongly increases at $T = 16$~K
and then drops again. Such behavior cannot be explained by the exciton-phonon
scattering because the rate of this scattering should monotonically depend 
on temperature in the range under study (see, e.g., Ref.~\cite{Poltavtsev}).
Possible physical mechanism of the exciton ejection will be discussed 
in the next section. 

Figure \ref{fig5}(b) shows the spectral dependence of differential
reflection measured at the time delay 30~ps when only the slow component
persists. The pump pulses were spectrally narrow as shown in the figure.
Although the spectral position of the pump pulses has been chosen to excite
predominantly the lowest exciton level, the non-zero signal is observed in
the wide spectral range up to the light-hole ($X_{lh}$) exciton transition.

\begin{figure}[t]
\includegraphics[width=1\columnwidth]{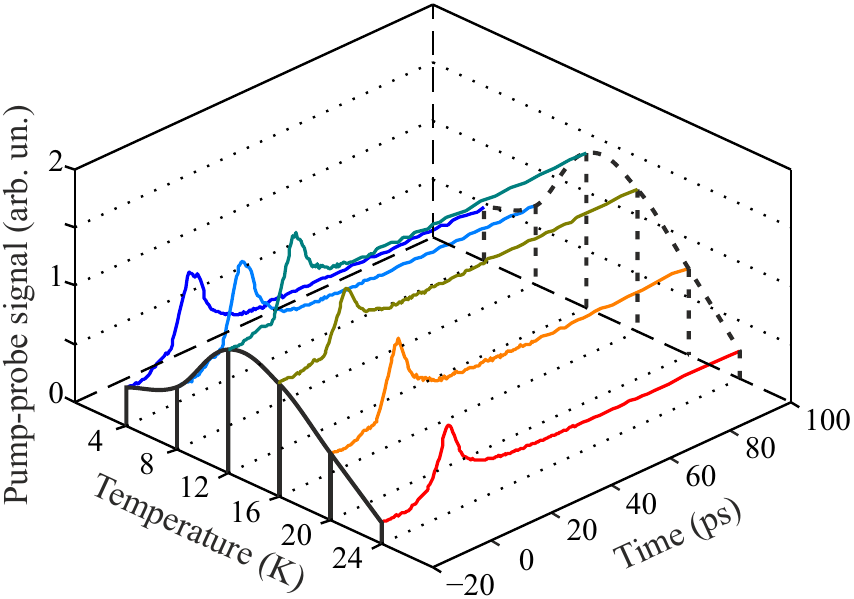}
\caption{(Color online) Temperature variations of the pump-probe signal for
the lowest exciton transition. $P_{exc} = 100$~$\protect\mu$W. }
\label{fig6}
\end{figure}

Temperature variations of the pump-probe signal are shown 
in Fig.~\ref{fig6}. One can see that a considerable signal appears at 
the negative delay where the probe pulse 
tests the sample before the pump pulse arrives. 
The temperature rise from 4~K to 12~K causes a considerable rise of this 
signal, which then drops at further increase of the temperature.

The presence of the pump-probe signal at negative delays indicates that the
characteristic relaxation time of the slow component, $\tau _{slow}$, exceeds
the repetition period of excitation pulses ($T_{l}=12.5$~ns). As a result,
the detected signal is accumulated from many preceding pump pulses. In the
case of exponential decay of the signal, the amplitude of the slow component
created by a single pulse becomes just before coming the next pulse equal to: 
$I_{pp1}=f_{slow}\exp (-T_{l}/\tau_{slow})$. The total signal for all the preceding 
pulses can be easily calculated as a sum of the geometric progression: 
$I_{pp}=f_{slow}\exp(-T_{l}/\tau _{slow})/[1-\exp (-T_{l}/\tau _{slow})]$. 
This expression allows us to roughly estimate the relaxation time $\tau _{slow}$ 
using the experimentally measured ratio $f_{slow}/I_{pp}$. 
Time  $\tau _{slow}$ increases from 15~ns to 45~ns in
the temperature range of $4-12$~K and rapidly drops with the further
temperature increase. At $T>24$~K, the signal at negative delays is not
detectable. This means that the decay time $\tau _{slow}$ is shorter than the
pulse repetition period $T_{l}$.

\section{Discussion}

\label{sect:disc}

The experimental results presented above clearly show that, at low
temperature and low excitation power, the structure under study is
characterized by extremely narrow widths of exciton resonances and
demonstrate zero Stokes shifts between the resonances observed in PL and
reflectance spectra. This is a clear indication of the absence of noticeable
inhomogeneous broadening of spectral lines. The line width is controlled
solely by the relaxation processes. 

The absence of any noticeable broadening of the resonances with the
temperature rise at quasiresonant excitation with low power (see Fig.~\ref{fig3})
means that the relaxation of radiative excitons is not affected by 
the exciton-phonon scattering in the temperature range studied. At the same
time, the peak width remarkably rises with excitation power (Fig.~\ref{fig4}). 
Because the pump power increase leads to the increase of the exciton
density, the observed additional broadening, $\delta E_c$, is most
probably related to exciton-exciton collisions resulting in the relaxation 
of exciton states. The rate of relaxation should be
proportional to the exciton density and, therefore, to the excitation power.
The experiment, however, shows a sublinear power dependence, see Fig.~4. 
We have to assume that the nonlinearity is caused by the collision-induced
decrease of the radiative recombination rate and of the corresponding
broadening, $\hbar\Gamma_0$, due to the decrease of the exciton coherence
volume. Such effect of additional relaxation on the radiative recombination rate
of excitons is theoretically discussed in Refs.~\cite{Hanamura-PRB1988,
Citrin-PRB1993, Ciuti-PRB1998} and experimentally studied in Refs.~%
\cite{Honold-PRB1989, Vinattieri-PRB1994, Alonso-PRB2003} The decrease of
radiative broadening partially compensates the collision-induced broadening so that
the total peak broadening, $\delta E=\hbar\Gamma_0 + \delta E_c$, should
depend on power sublinearly~\cite{Ciuti-PRB1998}. The linear dependence of
integral intensities of exciton peaks on the excitation power (not shown
here) evidences that there is no noticeable contribution of other
mechanisms, e.g., of nonradiative exciton recombination, to the line
broadening in these experimental conditions.

The above-noted increase of relative intensity and width of the first
exciton peak at high photon energy of excitation (see Fig.~\ref{fig2})
requires a particular attention. These observations point out that the
cascade relaxation of photocreated excitons over the quantum confined states
is replaced by the direct carrier relaxation into the lowest exciton level.
Indeed, at the excitation above the $X_{lh}$ transition, the probability of
resonant excitation with generation of excitons becomes smaller than that
of generation of free carriers. The photocreated electrons and
holes relax to their ground states where they are bound in excitons. The
excitons thus created populate predominantly the lowest energy state, that
explains the increased intensity of the $X1$ peak at these excitation conditions. 
Our assumption about the photocreation of free carriers is further supported 
by the observation of a strong broadening of the lowest exciton peak. 
Indeed, according to Refs.~\cite{Honold-PRB1989, Bajoni-PSS2006}, 
the cross-section of exciton-free-carrier scattering is by an order of 
magnitude larger than that of the exciton-exciton one. So, the
exciton relaxation due to exciton-carrier collisions is mainly
responsible for the strong broadening of the $X1$ resonance in the case 
of excitation above the $X_{lh}$ transition.

The temperature dependence of the PL peak intensities and widths (see 
Fig.~\ref{fig3}) appears to be contradictory at the first glance. The temperature
increase induces a remarkable decrease of the integral PL that points out to
the activation of efficient nonradiative relaxation processes for excitons.
The relaxation would seem to be accompanied by a noticeable broadening of
exciton peaks, which is not observed in the experiment.

\begin{figure}[t]
\includegraphics[width=0.7\columnwidth]{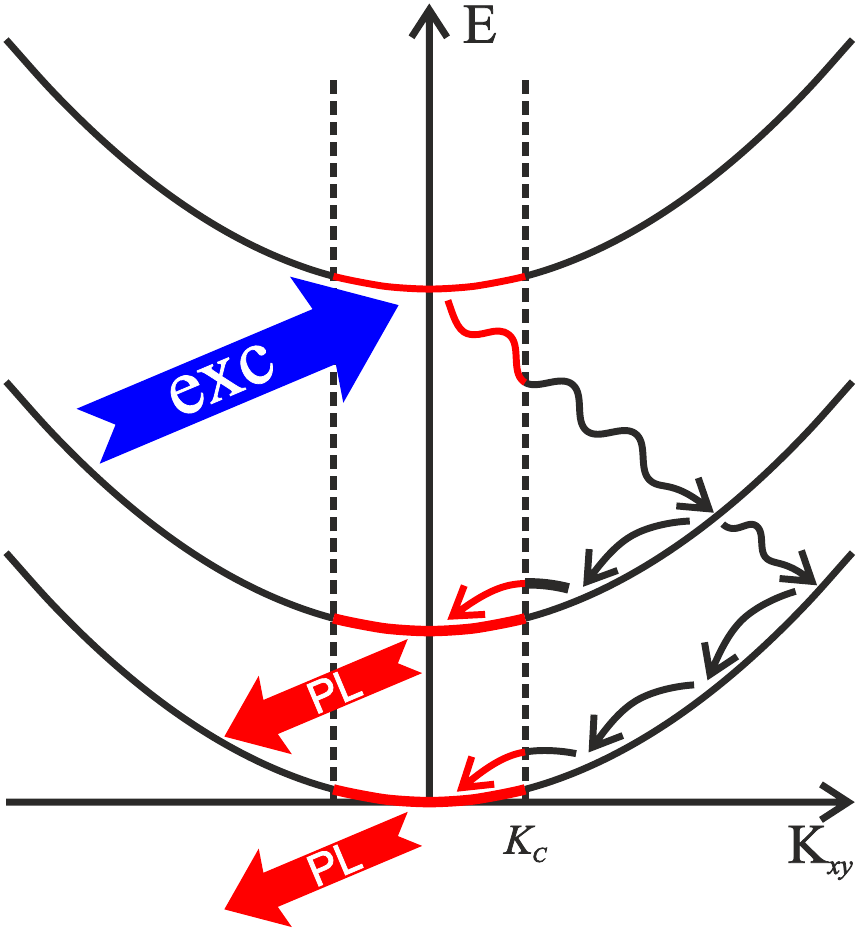}
\caption{(Color online) A simplified sketch of the processes discussed.
Parabolas show the dispersion of the quantum confined exciton states along 
the QW layer. Vertical dashed lines restrict the light cone area of wave
vectors. The bold arrows show excitation and PL processes and the wavy
arrows show the phonon-mediated exciton relaxation. }
\label{fig7}
\end{figure}

The origin of this effect is related to the temperature dependence of the
exciton density in the reservoir. The nonradiative excitons can be
efficiently created under nonresonant excitation via one-phonon relaxation
as schematically shown in Fig.~\ref{fig7}. According to the selection rule,
the exciton wave vector in the QW plane should to be equal to the projection
of the phonon wave vector onto the plane. The energy difference between the
nearest exciton levels in our structure is about 0.5 meV or larger. We can
estimate the wave vector acquired by an exciton in the course of relaxation
via emission of a longitudinal acoustic (LA) phonon. Note that the LA
phonons stronger interact with excitons than other acoustic phonons~%
\cite{Basu-PRB1992, Ivanov-PRB1999, Piermorocchi-PRB1996}. The wave 
vector of a LA phonon of energy about 0.5 meV in GaAs is 0.07 nm$^{-1}$, 
approximately. The light cone is limited by the photon wave vector, 
$K_{c}\approx $ 0.03 nm$^{-1}$ in GaAs, that is smaller than the phonon 
wave vector. So, the most part of excitons created by a nonresonant optical 
excitation should be nonradiative ones. 

At low temperature, the main decay channel of nonradiative excitons in
high-quality structure is the wave vector relaxation schematically shown in
Fig.~\ref{fig7}. As a result, nonradiative excitons convert into radiative
ones and eventually recombine. The rate of this relaxation should be
very limited because the radiative states are just a small fraction of all
states within the exciton band, and the probability for an exciton to be
found in one of the radiative states is low. This is why, the lifetime of
nonradiative excitons may drastically exceed that of radiative one.

We believe that nonradiative excitons are responsible for the long-lived
component of signal in pump-probe experiments, shown in Figs.~\ref{fig5}
and~\ref{fig6}. In these experiments, the reservoir of nonradiative excitons
can be filled in by ejection of radiative excitons as it is discussed in 
the preceding section. Besides, due to the finite spectral width of the pump 
pulses as well as the presence of some spectral wings [see Fig.~\ref{fig5}(b)],  
the excitation of the high-energy exciton transitions is also probable. 
Exciton relaxation from the excited states populates the reservoir 
of nonradiative excitons. The lifetime of
the long-lived component, $\tau _{slow}$, characterizes the rate of
transformation of the nonradiative excitons into radiative ones. At low
temperature $\tau _{slow}$ $\approx $15~ns, which is orders of 
magnitude longer than the lifetime of radiative excitons.

Nonradiative excitons do not directly interact with light. However, as we
can see, they give some indirect contribution to reflectance. Because of
orthogonality of the wave functions of excitons with different wave vectors,
their contribution to the pump-probe signal should not been connected with
bleaching of exciton transitions due to phase-space filling considered in
Refs~\cite{Schmitt-PRB1985,Hunsche-PRB2014,Koch-PRB1993}. The Coulomb screening effect discussed in these papers
is not expected to play a major role at low excitation powers used in our
experiments. More probably, the experimentally observed changes in
differential reflection are due to the scattering of radiative excitons by
nonradiative ones. The scattering results in the broadening of exciton
transitions, which is observed, particularly, in the PL spectra at strong
pumping, see Fig.~\ref{fig3}(d). The broadening changes the reflectivity,
which is detected as the pump-probe signal and, correspondingly, as the
change of the differential reflectance in the vicinity of high-energy exciton
transitions, observed experimentally, see Fig.~\ref{fig5}(b). 

The huge difference in lifetimes of the radiative and nonradiative excitons
explains the contradictory temperature behavior of the PL peaks mentioned
above. Indeed, the radiative states are mainly populated by the cascade
relaxation of photocreated excitons via nonradiative states. Due to the low
rate of transformation of nonradiative excitons into radiative ones,
almost all the excitation power is accumulated in the nonradiative exciton
reservoir. In particular, the ratio of densities of nonradiative and
radiative excitons, $n_{nr}/n_{r}=\tau _{slow}/T_1 \approx
2\times 10^{3}$ at $T=4$~K. The temperature increase triggers the exciton
dissociation primarily in this reservoir. The characteristic time of this
dissociation process becomes comparable to the lifetime of nonradiative
excitons which leads to an efficient depopulation of the reservoir. The
depopulation is evidenced by a rapid decrease of the slow component
amplitude in the pump-probe signal at $T>15$~K, see Fig.~\ref{fig5}(a). We
believe that this is the process responsible for the PL quenching at
elevated temperatures. At the same time, the exciton dissociation is still a
slow process in comparison with the radiative recombination and, therefore,
it cannot have a noticeable effect on the broadening of exciton peaks. This
explains the seeming contradiction mentioned above.

The interaction of radiative excitons with nonradiative ones explains the
nonmonotonic temperature dependence of the PL peak width. The increase of
the width when the temperature rises up to 15~K is related to accumulation
of nonradiative excitons, which is evidenced in the pump-probe experiments
as the increase of the amplitude of long-lived component of the signal (see
Fig.~6). The accumulation is caused by the increase of the average kinetic
energy and, consequently, of the average wave vector of the nonradiative
excitons with the temperature increase. As a result, the exciton relaxation
to the light cone is slowed down and their lifetime increases. 

\begin{figure}[t]
\includegraphics[width=1\columnwidth]{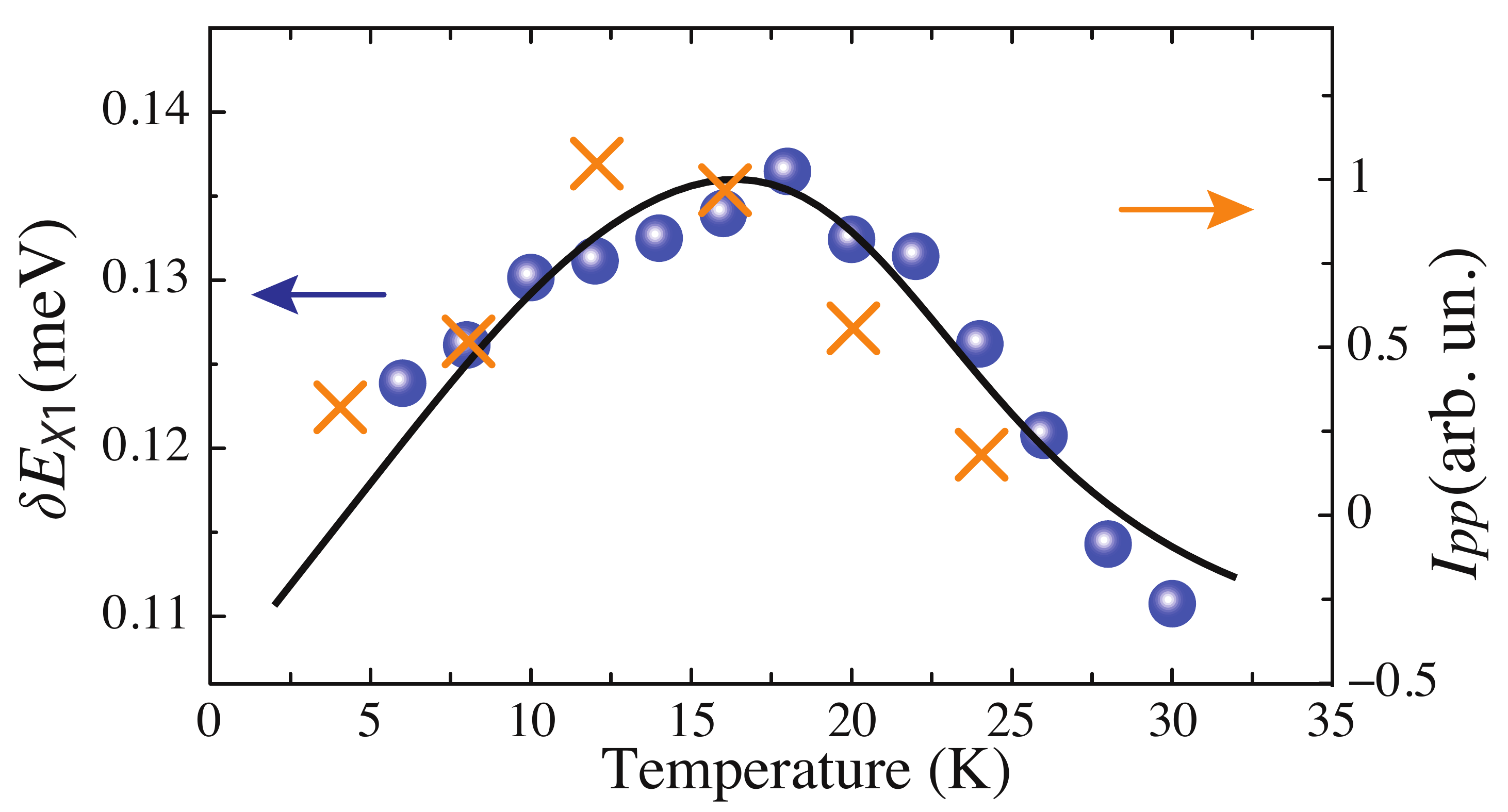}
\caption{(Color online) Temperature variations of the line broadening $\protect\delta E_{X1}$ 
(blue balls) for transition $X1$ measured at excitation power $P = 50$~$\protect\mu$W 
and of amplitude $I_{pp}$ (orange crosses) of the long-lived component 
of pump-probe signal shown in Fig.~\protect\ref{fig6}. Solid curve shows fit by Eq.~%
(\protect\ref{eq:14}) with parameters [see Eq.~(\protect\ref{eq:Arr})]: 
$\protect\delta E_0 = 120$~$\protect\mu$eV, $\protect\gamma_{X1}/\gamma_r = 14$, 
$\protect\gamma_b/\gamma_r=6000$. The activation energies are chosen 
the same as for PL quenching (see Fig.~\protect\ref{fig3}). }
\label{fig8}
\end{figure}

A simple theoretical analysis predicts a linear temperature dependence 
of the exciton lifetime and, hence, the density of nonradiative excitons in QWs~%
\cite{Andreani-SSC1991, Feldmann-PRL1987, Rosales-PRB2013, comment4}. At higher 
temperatures, the thermo-activated dissociation of nonradiative excitons reduces 
their density and, consequently, diminishes the additional broadening of exciton
peaks. Figure~\ref{fig8} shows that the peak width and the amplitude of the
long lived component exhibit likewise temperature dependencies for the first exciton
transition. This
confirms that nonradiative excitons strongly affect both broadening and
amplitude. The temperature dependence of $\delta E$ can be approximated by 
function:
\begin{equation}
\delta E_{X1}(T)=\delta E_{0}+\alpha T \times I(T),  
\label{eq:14}
\end{equation}
where $I(T)$ is given by Eq.~(\ref{eq:Arr}) and describes the thermally activated 
losses of excitons in the reservoir.

The nonmonotonic temperature behavior of the exciton ejection rate, 
$\gamma_{esc}$, discussed in Sect.~\ref{sect:temp-power} is also explained 
by the scattering of radiative excitons by the nonradiative ones. Rate 
$\gamma_{esc}$ normalized to the amplitude $I_{pp}$ of long-lived 
component of the pump-probe signal (see Fig.~\ref{fig8}) is found 
to be not dependent of temperature within an experimental error of about 15\%. 
Because $I_{pp}$ is proportional to the exciton density in the reservoir, 
$n_{nr}$, this fact means that $\gamma_{esc} \sim n_{nr}$. 
We, therefore, may conclude that the ejection of radiative excitons beyond 
the light cone is mainly caused by their collisions with nonradiative excitons. 

\section{Conclusion}

The analysis of PL spectra and relaxation kinetics of quantum confined exciton
states in a high-quality wide QW structure allowed us to obtain a valuable
information about the most important relaxation processes in this system.
Extremely narrow widths of spectral lines and the absence of any Stokes
shift between the resonances observed in the PL and reflectance spectra
indicate high quality of the structure. Our experiments show that, because
of the negligibly low density of defects, the most efficient process of
exciton decay at low temperatures and low excitation powers is the radiative
recombination. This process is responsible for the broadening of the low-energy 
exciton peak. The rate of radiative recombination for the higher
energy exciton states is reduced. This is why the phonon-mediated relaxation
of excitons from higher to lower energy states strongly affects the line
broadening. The relaxation process is additionally favored for higher energy
states since the density of acoustic phonon states increases as a function of energy.

Our study shows that the kinetics of radiative excitons is strongly affected
by the long-lived reservoir of nonradiative excitons whose wave vectors lie
outside the light cone. In particular, the reservoir manifests itself in the
long-lived component of the pump-probe signal. The lifetime of this
component is in three order of magnitude larger than the radiative decay
time. It is governed by slow relaxation of nonradiative excitons into the
radiative states. At low temperature, the slow relaxation does not lead to a
reduction of the PL yield, as nonradiative recombination processes in the
reservoir are inefficient in high-quality QWs. The temperature increase up
to 30~K results in the strong decrease of exciton PL intensity with no
noticeable broadening of the exciton lines. The origin of this unusual
effect is in the thermal dissociation of nonradiative excitons in the
reservoir. The dissociation rate is comparable to that of exciton relaxation
from nonradiative to radiative states but it is drastically lower than the
recombination rate of radiative excitons, which governs the widths of
exciton lines. The reservoir of nonradiative excitons also affects the
exciton PL spectra. In particular, the scattering of radiative excitons by
nonradiative ones at high pumping intensity results in a supplementary
broadening of the PL lines. 

In should be emphasized that the observed nontrivial effects are not 
specific property of the sample under study. They can be realized in any 
semiconductor QWs of moderate width characterized by ultra-high quality.

\section*{Acknowledgments}

The authors thank M. M. Glazov and K. V. Kavokin for fruitful discussions.
Financial support from the Russian Ministry of Science and Education
(contract no. 11.G34.31.0067), SPbU (grants No. 11.38.213.2014) is acknowledged. 
The authors also thank the SPbU Resource Center ``Nanophotonics'' 
(www.photon.spbu.ru) for the sample studied in present work. 
A.~V.~Kavokin acknowledges the support from EPSRC Established Career Fellowship.


\begin{thebibliography}{97}
\bibitem{Rashba} E. I. Rashba, G. E. Gurgenishvili, ``Edge absorption theory in semiconductors'',
Fiz. Tverd. Tela (Leningrad) \textbf{4}, 1029 (1962) [Sov. Phys. Solid State \textbf{4}, 759 (1962)].

\bibitem{Hanamura-PRB1988} Eiichi Hanamura, 
``Rapid radiative decay and enhanced optical nonlinearity of excitons in a quantum well'',  
\prb~\textbf{38}, 1228 (1988).

\bibitem{Deveaud-PRL1991} B. Deveaud, F. Cl\'erot, N. Roy, K. Satzke, B. Sermage, and D. S. Katzer, 
``Enhanced Radiative Recombination of Free Excitons in GaAs Quantum Wells'',
\prl~\textbf{67}, 2355 (1991).

\bibitem{Andreani-SSC1991} L.C. Andreani, F. Tassone, F. Bassani, 
``Radiative lifetime of free excitons in quantum wells'', 
Solid State Commun. \textbf{77}, 641, (1991).

\bibitem{Vinattieri-PRB1994} A. Vinattieri, Jagdeep Shah, T. C. Damen, D. S.
Kim, L. N. Pfeiffer, M. Z. Maialle, and L. J. Sham, 
``Exciton dynamics in GaAs quantum wells under resonant excitation'', 
\prb~\textbf{50}, 10868 (1994).

\bibitem{Koch-nmat2006} S. W. Koch, M. Kira, G. Khitrova, and H. M. Gibbs, 
``Semiconductor excitons in new light'', 
Nature Mat. \textbf{5}, 523 (2006).

\bibitem{Vishnevsky-PRB2012} D. V. Vishnevsky, D. D. Solnyshkov, N. A. Gippius, and G. Malpuech, 
``Multistability of cavity exciton polaritons affected by the thermally generated exciton reservoir'', 
\prb~\textbf{85}, 155328 (2012).

\bibitem{Wouters-PRB2013} M. Wouters, T. K. Para\"iso, Y. L\'eger, R. Cerna, F.
Morier-Genoud, M. T. Portella-Oberli, and B. Deveaud-Pl\'edran, 
``Influence of a nonradiative reservoir on polariton spin multistability'' 
\prb~\textbf{87}, 045303 (2013).

\bibitem{Giorgi-PRL2014} M. De Giorgi, D. Ballarini, P. Cazzato, G. Deligeorgis, S. I.
Tsintzos, Z. Hatzopoulos, P. G. Savvidis, G. Gigli, F. P. Laussy, and D. Sanvitto, 
``Temperature dependence of the radiative and nonradiative recombination time
in GaAs/Al$_x$Ga$_{1-x}$As quantum-well structures'', 
\prl~\textbf{112}, 113602 (2014).

\bibitem{Demirchyan-PRB2014} S. S. Demirchyan, I. Yu. Chestnov, A. P. Alodjants, M. M.
Glazov, and A. V. Kavokin, ``Qubits Based on Polariton Rabi Oscillators'', 
\prl~\textbf{112}, 196403 (2014).

\bibitem{Haug-PRB2014} H. Haug, T. D. Doan, and D. B. Tran Thoai, 
``Quantum kinetic derivation of the nonequilibrium Gross-Pitaevskii equation for nonresonant excitation of microcavity polaritons'', 
\prb~\textbf{89}, 155302 (2014).

\bibitem{Belykh-PRB2014} V. V. Belykh and D. N. Sobyanin, 
``Polariton linewidth and the reservoir temperature dynamics in a semiconductor microcavity'', 
\prb~\textbf{89}, 245312 (2014).

\bibitem{Piermorocchi-PRB1996} C. Piermarocchi, F. Tassone, V. Savona, A.
Quattropani, and P. Schwendimann, ``Nonequilibrium dynamics of free quantum-well excitons in time-resolved photoluminescence'', 
\prb~\textbf{53}, 15834 (1996).

\bibitem{Feldmann-PRL1987} J. Feldmann, G. Peter, E. O. G\''obel, P. Dawson, K.
Moore, C. Foxon, and R. J. Elliott, ``Linewidth Dependence of Radiative Exciton Lifetimes in Quantum Wells'', 
\prl~\textbf{59}, 2337 (1987).

\bibitem{Citrin-PRB1993} D. S. Citrin, 
``Radiative lifetimes of excitons in quantum wells: Localization and phase-coherence effects'',  
\prb~\textbf{47}, 3832 (1993).

\bibitem{Kavokin-PRB1994} A. V. Kavokin, 
``Exciton oscillator strength in quantum wells: From localized to free resonant states'', 
\prb~\textbf{50}, 8000 (1994).

\bibitem{Damen-PRB1990} T. C. Damen, Jagdeep Shah, D. Y. Oberli, D. S. Chemla, J. E. Cunningham, and J. M. Kuo, 
``Dynamics of exciton formation and relaxation in GaAs quantum wells'', 
\prb~\textbf{42}, 7434 (1990).

\bibitem{Srinivas-PRB1992} Vivek Srinivas, John Hryniewicz, Yung Jui Chen, and Colin E. C. Wood, 
``Intrinsic linewidths and radiative lifetimes of free excitons in GaAs quantum wells'',
\prb~\textbf{46}, 10193 (1992).

\bibitem{Szczytko-PRL2004} J. Szczytko, L. Kappei, J. Berney, F. Morier-Genoud, M.T. Portella-Oberli, and B. Deveaud, 
``Determination of the Exciton Formation in Quantum Wells from Time-Resolved Interband Luminescence'', 
\prl~\textbf{93}, 137401 (2004).

\bibitem{Roussignol-PRB1992} Ph. Roussignol, C. Delalande, A. Vinattieri, L. Carraresi, and M. Colocci, 
``Dynamics of exciton relaxation in GaAs/Al$_x$Ga$_{1-x}$As quantum wells'',  
\prb~\textbf{45}, 6965 (1992).

\bibitem{Kappei-PRL2005} L. Kappei, J. Szczytko, F. Morier-Genoud, and B. Deveaud, 
``Direct Observation of the Mott Transition in an Optically Excited Semiconductor Quantum Well'', 
\prl~\textbf{94}, 147403 (2005).

\bibitem{Deveaud-ChemPhys2005} B. Deveaud, L. Kappei, J. Berney, F. Morier-Genoud, M.T. Portella-Oberli, J. Szczytko, and C. Piermarocchi, 
``Excitonic effects in the luminescence of quantum wells'', 
Chem. Phys. \textbf{318}, 104 (2005).

\bibitem{Bajoni-PSS2006} D. Bajoni, P. Senellart, M. Perrin, A. Lemai\^tre, B. Sermage, and J. Bloch, 
``Exciton dynamics in the presence of an electron gas in GaAs quantum wells'',  
phys. stat. sol. (b) \textbf{243}, 2384 (2006).

\bibitem{Portella-PRL2009} M. T. Portella-Oberli, J. Berney, L. Kappei, F. Morier-Genoud, J. Szczytko, and B. Deveaud-Pl\'edran, 
``Dynamics of Trion Formation in In$_x$Ga$_{1-x}$As Quantum Wells'', 
\prl~\textbf{102}, 096402 (2009).

\bibitem{Basu-PRB1992} P. K. Basu and Partha Ray, 
``Energy relaxation of hot two-dimensional excitons in a GaAs quantum well by exciton-phonon interaction'', 
\prb~\textbf{45}, 1907 (1992).

\bibitem{Ciuti-PRB1998} C. Ciuti, V. Savona, C. Piermarocchi, A. Quattropani, and P. Schwendimann, 
``Role of the exchange of carriers in elastic exciton-exciton scattering in quantum wells'', 
\prb~\textbf{58}, 7926 (1998).

\bibitem{Ivanov-PRB1999} A. L. Ivanov, P. B. Littlewood, and H. Haug, 
``Bose-Einstein statistics in thermalization and photoluminescence of quantum-well excitons'', 
\prb~\textbf{59}, 5032 (1999).

\bibitem{Piermarocchi-pss1997} C. Piermarocchi, V. Savona, A. Quattropani, P. Schwendimann, and F. Tassone, 
``Photoluminescence and Carrier Dynamics in GaAs Quantum Wells'',
 phys. stat. sol. (a) \textbf{164}, 221 (1997).

\bibitem{Gurioli-PRB1991} M. Gurioli, A. Vinattieri, M. Colocci, C. Deparis, J. Massies, G. Neu, A. Bosacchi, and S. Franchi, 
``Temperature dependence of the radiative and nonradiative recombination time in GaAs/Al$_x$Ga$_{1-x}$As quantum-well structures'',  
\prb~\textbf{44}, 3115 (1991).

\bibitem{Eccleston-PRB1992} R. Eccleston, B. F. Feuerbacher, J. Kuhl, W. W. R\"uhle, and K. Ploog, 
``Density-dependent exciton radiative lifetimes in GaAs quantum wells'', 
\prb~\textbf{45}, 11403 (1992).

\bibitem{Filkenstein-PRB1998} G. Finkelstein, V. Umansky, I. Bar-Joseph, V. Ciulin, S. Haacke, J.-D. Gani\'ere, and B. Deveaud, 
``Charged exciton dynamics in GaAs quantum wells'', 
\prb~\textbf{58}, 12637 (1998).

\bibitem{Schultheis-PRB1986} L. Schultheis, A. Honold, J. Kuhl, K. K\"ohler, and C. W. Tu, 
``Optical dephasing of homogeneously broadened two-dimensional exciton transitions in GaAs quantum wells'', 
\prb~\textbf{34}, 9027 (1986).

\bibitem{Honold-PRB1989} A. Honold, L. Schultheis, J. Kuhl, and C. W. Tu, 
``Collision broadening of two-dimensional excitons in a GaAs single quantum well'', 
\prb~\textbf{40}, 6442 (1989).

\bibitem{Kim-PRL1992} Dai-Sik Kim, Jagdeep Shah, T. C. Damen, W. Sch\"afer, F. Jahnke, S. Schmitt-Rink, and K. K\"ohler, 
``Unusually Slow Temporal Evolution of Femtosecond Four-Wave-Mixing Signals in Intrinsic GaAs Quantum Wells: 
Direct Evidence for the Dominance of Interaction Effects'', 
\prl~\textbf{69}, 2725 (1992).

\bibitem{Duer-PRL1997} R. Duer, I. Shtrichman, D. Gershoni, and E. Ehrenfreund, 
``Momentum Redistribution Times of Resonantly Photogenerated 2D Excitons'', 
\prl~\textbf{78}, 3919 (1997).

\bibitem{Borri-PRB1999} P. Borri, W. Langbein, J. M. Hvam, and F. Martelli, 
``Well-width dependence of exciton-phonon scattering in In$_x$Ga$_{1-x}$As/GaAs single quantum wells'', 
\prb~\textbf{59}, 2215 (1999).

\bibitem{Chemla-Nature2001} D. S. Chemla and Jagdeep Shah, 
``Many-body and correlation effects in semiconductors'', 
Nature \textbf{411}, 549 (2001).

\bibitem{Smith-PRL2010} R. P. Smith, J. K. Wahlstrand, A. C. Funk, R. P.
Mirin, S. T. Cundiff, J. T. Steiner, M. Schafer, M. Kira, and S.W. Koch, 
``Extraction of Many-Body Configurations from Nonlinear Absorption in Semiconductor Quantum Wells'', 
\prl~\textbf{104}, 247401 (2010).

\bibitem{Uddin-JAP1989} A. Uddin and T. G. Andersson, 
``Investigation of high-quality GaAs:In layers grown by molecular-beam epitaxy'', 
J. Appl. Phys. \textbf{65}, 3101 (1989).

\bibitem{Ivchenko-book} E. L. Ivchenko, ``Optical spectroscopy of semiconductor nanostructures'', 
Springer (Berlin) 2004, 437 p.

\bibitem{Loginov} E. V. Ubyivovk, D. K. Loginov, I. Ya. Gerlovin, Yu. K. Dolgikh, Yu. P. Efimov, 
S. A. Eliseev, V. V. Petrov, O. F. Vyvenko, A. A. Sitnikova, and D. A. Kirilenko, 
``Experimental determination of dead layer thickness for excitons in a wide GaAs/AlGaAs quantum well'',  
Fiz. Tv. Tela \textbf{51}, 1818 (2009) [Phys. Sol. State \textbf{51}, 1929 (2009)].

\bibitem{Schiumarini-PRB2010} D. Schiumarini, N. Tomassini, L. Pilozzi, and A.
D'Andrea, ``Polariton propagation in weak-confinement quantum wells',  \prb~\textbf{82}, 075303 (2010).

\bibitem{TredicuchiPRB1993} A. Tredicucci, Y. Chen, F. Bassani, J. Massies, C. Deparis, G. Neu, 
``Center-of-mass quantization of excitons and polariton interference in GaAs thin layers'', 
\prb~\textbf{47}, 10348 (1993).

\bibitem{Muraki-APL1992} K. Muraki, S. Fukatsu, and Y. Shiraki, 
``Surface segregation of In atoms during molecular beam epitaxy and 
its influence on the energy levels in InGaAs/GaAs quantum wells'', 
Appl. Phys. Lett. {\bf 61}, 557 (1992).

\bibitem{Poltavtsev} S.V. Poltavtsev, Yu.P. Efimov, Yu.K. Dolgikh, S.A. Eliseev, V.V. Petrov, V.V. Ovsyankin, 
``Extremely low inhomogeneous broadening of exciton lines in shallow (In,Ga)As/GaAs quantum wells'',  
Solid State Commun.~\textbf{199}, 47 (2014)

\bibitem{Chen87} Y.J. Chen, Emil S. Koteles, Johnson Lee, J.Y. Chi, B.S. Elman, 
``Temperature dependent optical studies of GaAs/AlGaAs single quantum wells'', 
SPIE~\textbf{792}, 162 (1987).

\bibitem{Singh-PRB13} Rohan Singh, Travis M. Autry, Ga\"el Nardin, Galan
Moody, Hebin Li, Klaus Pierz, Mark Bieler, and Steven T. Cundiff, 
``Anisotropic homogeneous linewidth of the heavy-hole exciton in (110)-oriented GaAs quantum wells'', 
\prb~\textbf{88}, 045304 (2013).

\bibitem{Tuffigo} H. Tuffigo, R. T. Cox, N. Magnea, Y. Merle d'Aubign\'e, and A. Million, 
``Luminescence from quantized exciton-polariton states in Cd$_{1-x}$Zn$_x$Te/CdTe/Cd$_{1-x}$Zn$_x$Te
thin-layer heterostructures'',  
\prb~\textbf{37}, 4310 (1988).

\bibitem{footnote} The low magnitude quasiperiodic oscillations of the PL
intensity are due to the interference of the excitation laser light
reflected from the front and rear surfaces of the sample.

\bibitem{NCP} E.L. Ivchenko, G.E. Pikus, B.S. Razbirin, A.I. Starukhin, 
``Optical orientation and alignment of free excitons in GaSe during resonant excitation - Theory'', 
Russian ZhETF \textbf{72}, 2230 (1977) [JETP \textbf{45}, 1172 (1977)].

\bibitem{Rosales-PRB2013} D. Rosales, T. Bretagnon, B. Gil, A. Kahouli, J. Brault, 
B. Damilano, J. Massies, M. V. Durnev, and A. V. Kavokin, 
``Excitons in nitride heterostructures: From zero- to one-dimensional behavior'',  
Phys. Rev. B \textbf{88}, 125437(2013).

\bibitem{JoycePRB91} M. J. Joyce, Z. Y. Xu, and M. Gal, 
``Photoluminescence excitation spectroscopy of as-grown and chemically released In$_0.05$Ga$_0.95$As/GaAs quantum wells'', 
\prb~\textbf{44}, 3144 (1991).

\bibitem{ZubkovPRB04} V. I. Zubkov, M. A. Melnik, A. V. Solomonov, and E. O. Tsvelev, F. Bugge, M. Weyers, and G. Tr\"ankle, 
``Determination of band offsets in strained In$_x$Ga$_{1-x}$As/GaAs quantum wells 
by capacitance-voltage profiling and Schr\''odinger-Poisson self-consistent simulation'',  
\prb~\textbf{70}, 075312 (2004).

\bibitem{Biswas-Materials14} Dipankar Biswas, Partha Pratim Bera, Tapas Das, Siddhartha Panda, 
``Impact of strain on the band offsets of important III-V quantum wells: 
In$_x$Ga$_{1-x}$N/GaN, GaAs/In$_x$Ga$_{1-x}$P and In$_x$Ga$_{1-x}$As/AlGaAs'',  
Materials Sci. Semicond. Processing~\textbf{19}, 11 (2014).

\bibitem{Shah-book} J. Shah, 
``Ultrafast spectroscopy of semiconductor and semiconductor nanostructures'', 
Berlin, ``Springer-Verlag'', 1996, 372 pp.

\bibitem{Schmitt-PRB1985} S. Schmitt-Rink, D. S. Chemla, and D. A. B. Miller, 
``Theory of transient excitonic optical nonlinearities in semiconductor quantum-well structures'', 
\prb~\textbf{32}, 6601 (1985).

\bibitem{Hunsche-PRB2014} S. Hunsche, K. Leo, H. Kurz, and K. K\"ohler, 
``Exciton absorption saturation by phase-space filling: Influence of carrier temperature and density'', 
\prb~\textbf{49}, 16565 (1994).

\bibitem{Koch-PRB1993} M. Koch, J. Feldmann, E.O. G\"obel, P. Thomas, and J. Shah, and K. K\"ohler, 
``Coupling of exciton transitions associated with different quantum-well islands'',  
\prb~\textbf{48}, 11480 (1993).

\bibitem{Alonso-PRB2003} F\'elix Fern\'andez-Alonso, Marcofabio Righini, Andrea Franco, and Stefano Selci, 
``Time-resolved differential reflectivity as a probe of on-resonance exciton dynamics in quantum wells'', 
\prb~\textbf{67}, 165328 (2003).

\bibitem{comment4} In our case of relatively wide QW, the density of states
of nonradiative excitons is a step function with steps at the each
quantum confined state. Therefore, the temperature dependence of the exciton
lifetime deviates from the linear one proposed for two dimensional excitons
and tends to approach the $T^{3/2}$ dependence characteristic of the bulk
material, see, e.g., Ref.~\protect\cite{Rosales-PRB2013}. We use the linear 
dependence in the fit in Fig.~\ref{fig8} in order
to minimize the number of fitting parameters.

\end{thebibliography}
\end{document}